\documentclass[aps,epsf,preprint]{revtex4}

\def\1{{\bf 1}}
\def\[{\left[}
\def\]{\right]}
\def\be{\begin{eqnarray}}
\def\ee{\end{eqnarray}}

\def\({\left(}
\def\){\right)}
\def\bk#1{\langle#1\rangle}
\def\eq#1{(\ref{#1})}

\def\o{\omega}

\def\f{\phi}
\def\q{\psi}
\def\G{{\cal G}}
\def\l{\lambda}
\def\m{\mu}

\begin{document}

\title{The Horizontal Symmetry for Neutrino Mixing}
\author{C.S. Lam}
\address{Department of Physics, McGill University\\
 Montreal, Q.C., Canada H3A 2T8\\
and\\
Department of Physics and Astronomy, University of British Columbia,  Vancouver, BC, Canada V6T 1Z1 \\
Email: Lam@physics.mcgill.ca}
%\date{\today}

\begin{abstract}
We argue that the best way to determine horizontal symmetry is from neutrino mixing, and proceed to show that the only 
finite group capable of yielding the tri-bimaximal mixing for all Yukawa couplings is $S_4$, or any group containing it.
The method used is largely group theoretical, but it can be implemented by dynamical schemes in which  
the Higgs expectation values introduced to break $S_4$ spontaneously are uniquely
determined up to an unknown scale for each. 
 
\end{abstract}
%\pacs{}
\narrowtext
\maketitle

\section{Introduction}
Progress in particle physic is often guided by symmetry. From isospin to the eightfoldway, from the Standard Model (SM) to GUT, SUSY and superstrings, symmetry always plays a central role. It is therefore natural to expect that symmetry
may open the door to the generation problem as well. 
For that reason, a plethora of  horizontal symmetry groups have been proposed, 
including $Z_m$, $Z_m\times
Z_n,\ D_n,\ S_3,\ S_4,\ A_4,\ A_5,\ T',\ \Delta(27), SO(3), SU(3)$, and others. The reason why so many diverse groups 
can all claim to be reasonable candidates is the presence of many adjustable Yukawa coupling constants and Higgs expectation
values in these models. By suitably tuning these parameters one
can arrive at many attractive results.

If there is indeed a horizontal symmetry in nature, it must be unique, and we need a criterion to determine what it is.
I subscribe to the view that a true symmetry would reveal itself without any tuning of the dynamical parameters,
and I shall use that as the criterion to determine the horizontal symmetry.
I argue that neutrino mixing, rather than quark mixing or
the fermion masses, is the proper vehicle to fix such a symmetry.
This latter assertion may be contrary to the instinct built up from atomic physics,
where approximate symmetry is reflected by proximity of energy levels.
In particle physics, symmetries are often broken spontaneously by a large extent, 
rendering mass spectra useless for recovering the unbroken symmetry. For example, in SM, the bottom
and the top quarks belong to the same isodoublet, but their masses are so vastly different that no trace
is left of the isospin symmetry. Similarly, the masses of the quarks 
and charged leptons in different generations
are also very different, suggesting that horizontal symmetry is also broken spontaneously
and fermion masses are useless in its recovery. I also think that 
quark mixing, being small, may result from a complicated dynamical perturbation of the unmixed state,
whereas neutrino mixing, being large and regular, can best be used to find out
 the unperturbed and the unbroken horizontal symmetry.
The regularity of tri-bimaximal mixing \cite{HPS} of neutrinos is analogous to
 the regularity of the Balmer series for hydrogen
atom. The latter led to the discovery of the Bohr atom, with its rotational and dynamical symmetry
of a Coulomb potential, but it cannot predict fine structures and hyperfine structures of the spectra 
brought on by additional dynamical
perturbations.
Similarly, the tri-bimaximal mixing may also be subject to a small perturbation which future experiments 
will reveal,
but that does not invalidate the horizontal symmetry established by its use.

I shall show in this letter and a subsequent detailed paper \cite{LAM2} that  
$S_4$, the permutation group of four objects and the symmetry group of the octahedron and the cube, is the only finite
group capable of giving rise to tri-bimaximal mixing without tuning parameters. 
This symmetry is {\it unique} up to the obvious generalization, that any group containing $S_4$
is a possible horizontal group as well. To avoid repetition, when we say $S_4$ is unique from now on,
we always mean to include this possible extension.

Since we like to uncover the symmetry without resorting to specific dynamics, 
the method employed is largely group theoretical, but we will discuss the implementation of some dynamical
schemes at the end. In that case,  $S_4$ is broken by the introduction of Yukawa couplings and non-SM Higgs bosons.
The Higgs expectation values are uniquely
determined by the group structure, up to unknown scales that will be absorbed into the Yukawa coupling constants
to form `effective coupling constants', to be
used to fit the leptonic masses. Since there are now additional Higgs present to share the burden of 
fermion masses, the coupling of the SM Higgs to leptons are {\it no longer} proportional to their masses.

Much has been written about the $S_4$ subgroup $A_4$ as a horizontal group \cite{A4}. However, $A_4$ gives rise naturally
only to trimaximal mixing but not bimaximal mixing \cite{LAM}. It requires either a tuning of the Yukawa couplings 
\cite{MA}, or the additional symmetries contained in $S_4$ to get the bimaximal mixing. The group $S_4$ had 
been previously studied \cite{S4}, but with a different motivation and a different conclusion.

\section{From tri-bimaximal mixing to $S_4$}
After reviewing \cite{LAM} how $S_4$ comes about, the argument for its uniqueness will be outlined.

Let $c=(e_L,\mu_L,\tau_L)^T$ be the left-handed charged leptons and $\nu=(\nu_e,\nu_\mu,\nu_\tau)^T$  the left-handed
Majorana neutrinos. Instead of their mass matrices $M_c$ and $M_\nu$, 
we study the combination $\bar M_c=\sqrt{M_cM_c^\dagger}$ and $M_\nu$, because they connect
left-handed to left-handed fermions, thereby avoiding
 the involvement of the right-handed fermions in this symmetry
analysis. $\bar M_c$ is hermitean and $M_\nu$ symmetric; they can be diagonalized
by unitary matrices $U_c$ and $U_\nu$, so that $U_c^\dagger \bar M_c U_c$ is the diagonal matrix of charge-lepton
masses, and $U_\nu^TM_\nu U_\nu$ is the diagonal matrix of neutrino masses. The PMNS
mixing matrix is given by $U=U_c^\dagger U_\nu$. If $F$ is a symmetry operation of $c$ and $G$ a symmetry operation
 of $\nu$, both
unitary, then under the transformations $c\to Fc$ and $\nu\to G\nu$, symmetry demands
$F^\dagger\bar M_cF=\bar M_c$ and 
$G^TM_\nu G=M_\nu$. As shown in \cite{LAM}, this means that the eigenvectors of $F$ are
the columns of $U_c$, with eigenvalues of unit modulus, and the eigenvectors
of $G$ are the columns of $U_\nu$, with eigenvalues $\pm 1$.
We shall choose the sign of $G$ so that it has one $+1$ eigenvalue and two $-1$'s.. 

It follows \cite{LAM} that if $F=G$, then $U_c=U_\nu$ and $U={\bf 1}$. 
This is false, hence the horizontal symmetry must be
broken to enable $F\not= G$, and we assume the breaking to be spontaneous. 
In the basis where $M_c$ is diagonal, which we
adopt from now on, $F$ is diagonal and $U=U_\nu$. Hence the neutrino symmetry operator $G$ can be read off
from the tri-bimaximal mixing matrix $U$. There are three of them, with 
the eigenvector of $G_i\ (i=1,2,3)$ of eigenvalue $+1$ taken from the $i$th column of $U$, and the
other two eigenvectors of eigenvalues $-1$ taken from the other two columns. See \cite{LAM} for details and formulas.
These three matrices commute, with $G_1=G_2G_3$, so the group containing $G_2$ and $G_3$ must also 
automatically contain $G_1$. The
minimal horizontal group appropriate to tri-bimaximal mixing is therefore the finite group ${\cal G}=\{F,G_2,G_3\}$ generated
by $F, G_2$, and $G_3$. This group is not a priori unique because $F$ is not. However, since
${\cal G}$ is assumed to be finite, 
there must be an integer $n$ such that $F^n={\bf 1}$. Conversely, given a finite group $\G$,
it can be spontaneously broken to reveal the tri-bimaximal mixing without tuning only when three of its members, 
$F,G_2,G_3$, can be
found to have these properties when $\bar M_c$ is diagonal. 
Since $\bar M_c$ is not known from $\G$, the only way
to guarantee its diagonality is to go to the basis where $F$ is diagonal. Since $F$ commutes with $\bar M_c$,
the diagonality of $\bar M_c$ is guaranteed
if the three eigenvalues of $F$ are
 different, so we shall demand that of $F$ from now on. In particular, this requires
$n\ge 3$. 
For $n=3$, the three entries of $F$ must be 1, 
$\o=\exp(2\pi i/3)$, and $\o^2$. There are $3!=6$ possible $F$'s obtained from different positioning of these three
eigenvalues, 
but they only generate two different groups, ${\cal G}=
S_4$, and $3.S_4$ \cite{FN}. The latter is obtained by adjoining $S_4$ with $\o S_4$ and $\o^2 S_4$, and it contains
$S_4$ as a subgroup. So for $n=3$, the minimum horizontal group is $S_4$.

To prove the uniqueness of $S_4$, we must show that no other finite group (except those containing $S_4$) can be
so generated for $n>3$.
A direct proof is difficult because there are an infinite number of cases to consider, so we shall resort to
 a different strategy. Since  an overall scalar
factor multiplying a matrix does not alter its eigenvectors, which are all that we care in order to get the correct $U$, 
we may confine ourselves to finite subgroups of $SU(3)$ and $SO(3)$, 
or their central extensions. 
We must show that unless the finite group contains $S_4$, otherwise it is impossible to find three members $F,G_2,G_3$ 
in it so that in the basis where $F$ is diagonal, the invariant eigenvectors of $G_2$ and $G_3$ 
are given by the second and third columns of the 
tri-bimaximal matrix $U$. This strategy is more viable than a direct approach
because all the finite subgroups
of $SO(3)$ (or $SU(2)$) and $SU(3)$ are known.

For $SO(3)$ (or $SU(2)$) \cite{SU2}, they are given by the two infinite series, $Z_n$ (cyclic groups)
and $D_n$ (dihedral groups), and  three isolated ones: $A_4$, the alternating group
of 4 objects, which is also the symmetry group of the tetrahedron; 
$S_4$, the symmetric group of four objects,  which is also the symmetry group of the octahedron and the cube; 
and $A_5$, the symmetry group of the icosahedron and the dodecahedron. For $SU(3)$ \cite{SU2, SU3}, 
there are again two infinite series, $\Delta(3n^2)$
and $\Delta(6n^2)$, and six isolated ones, $\Sigma(36), \Sigma(60), \Sigma(72), \Sigma(168), \Sigma(216)$, 
and $\Sigma(360)$; the number in each case indicates the order of the group. 
The detailed argument to reject all of them except $S_4$ is somewhat lengthy, and will be postponed
to another publication \cite{LAM2}. However, it is easy to state on what basis each of them is rejected. First of all,
it has been shown in \cite{LAM} that the group must possess a three-dimensional irreducible representation, or else
we cannot get the tri-bimaximal mixing pattern without tuning parameters. 
On that basis the groups $Z_n, D_n, \Sigma(36), \Sigma(72),  
\Sigma(360)$ can be rejected because they do not possess
three-dimensional irreducible representations. As mentioned before, the group $A_4$ is rejected because it leads to
trimaximal but not bimaximal mixing \cite{LAM}. The groups $\Delta(3n^2)$ and $\Delta(6n^2)$ are rejected because
their explicitly-known three-dimensional irreducible representations all have a special form, so special
 that tri-bimaximal mixing cannot occur unless they contain $S_4$ as a subgroup.
The rest of the groups are rejected by using their character tables to pick out the order $n$ of $F$ and its eigenvalues.
If $n=3$, then either the group contains $S_4$ or else it cannot accommodate the tri-bimaximal mixing. For $n>3$,
we can use its eigenvalues to construct  all possible $F$. With $G_2, G_3$ determined from the columns of the
tri-bimaximal matrix, we can compute the orders of $FG_2$ and $FG_3$,
and in each case one or the other would have an order larger than the order of the whole
finite group. Hence at least one of these $G_2$ and $G_3$ cannot be in the group and tri-bimaximal mixing cannot occur.

\section{Spontaneous breaking}
The discussion so far is purely group theoretical. To implement a dynamical scheme complementary
to the discussion we have to write down 
the mass term of an effective Hamiltonian. After integrating over the right-handed fermions,
it can be symbolically written as
\be
H=\sum_A\(\lambda_A c^\dagger c \phi^A+\mu_A \nu^T\nu\psi^A\)+h.c.,\label{H} \ee
where $\lambda_A$ and $\mu_A$ are the Yukawa coupling constants to the Higgs fields $\phi^A$ and $\psi^A$. For later
convenience, an energy scale is incorporated into the couplings
so that the Higgs fields become dimensionless, with vacuum expectation
values given later. The Higgs fields in \eq{H} may be composite, and
the spacetime
structure is implicit, may even be non-local, but all that we care is the $S_4$ behavior. Before the Higgs bosons develop an
expectation value, $H$ must be invariant under every $S_4$ transformation. Afterwards, the horizontal symmetry
is broken, $\bk{H}$ is no longer invariant under every $S_4$, but it must still be invariant under the residual symmetries
of $F$ on $c$, and $G_2, G_3$ on $\nu$, in order to recover the tri-bimaximal mixing. To achieve that, we must have 
\be 
F\bk{\phi^A}=\bk{\phi^A},\ G_{2,3}\bk{\psi^A}=\bk{\psi^A}.\label{FG} \ee
These equations determine the structure of the vacuum expectation value for every Higgs boson up to an unknown scale
which has been incorporated into the Yukawa couplings.

$S_4$ has five irreducible
representations, ${\bf 1, 1', 2, 3, 3'}$, and by definition
the left-handed fermions belong to ${\bf 3}$. If we use a boldface superscript to denote an irreducible representation, then the representations of $F$ and $G_i\ (i=2,3)$ are:
$F^{\bf 1}= F^{\bf 1'}=G_i^{\bf 1}=G_2^{\bf 1'}=-G_3^{\bf 1'}=1,\
F^{\bf 2}={\rm diag}(\o,\o^2),\ 
F^{\bf 3}=F^{\bf 3'}={\rm diag}(1,\o,\o^2),\ G_2^{\bf 2}={\rm diag}(1,1)$, and
\be G_3^{\bf 2}=\pmatrix{0&1\cr 1&0\cr},\
G_2^{\bf 3}=G_2^{\bf 3'}={1\over 3}\pmatrix{-1&2&2\cr 2&-1&2\cr 2&2&-1},\ 
G_3^{\bf 3}=-G_3^{\bf 3'}=-\pmatrix{1&0&0\cr 0&0&1\cr 0&1&0\cr}.\ee
Applying this to \eq{FG}, we deduce that $\bk{\f^{\bf 1}}=\bk{\f^{\bf 1'}}=\bk{\q^{\bf 1}}=1$,
$\bk{\q^{\bf 1'}}=\bk{\q^{\bf 3}}=\bk{\f^{\bf 2}}=0$, 
$\bk{\f^{\bf 3}}=\bk{\f^{\bf 3'}}=(1,0,0)^T$, $\bk{\psi^{\bf 2}}=(1,1)^T$, and $\bk{\q^{\bf 3'}}=(1,1,1)^T$. 
Since ${\bf 3}\times {\bf 3}$ produces ${\bf 1}+{\bf 2}+{\bf 3}+{\bf 3'}$, neither $\f^{\bf 1'}$ nor $\q^{\bf 1'}$
is present in \eq{H}. With $\bk{\f^{\bf 2}}=\bk{\q^{\bf 3}}=0$, there remain exactly three Yukawa
coupling constants each for the charged leptons and neutrinos in $\bk{H}$, just enough to fit the three charged lepton
masses and the three neutrino masses. With appropriate Clebsch-Gordan coefficients inserted, the mass matrices can
be read off from \eq{H} to be $\bar M_c={\rm diag}(a-2b,a+b-c,a+b+c)$, where $a=\l^{\bf 1}/\sqrt{3}, b=\l^{\bf 3'}/\sqrt{6}, 
c=\l^{\bf 3}/\sqrt{2}$, and
\be
M_\nu=\pmatrix{c-2e&d+e&d+e\cr d+e&d-2e&c+e\cr d+e&c+e&d-2e\cr},\ee
where $d=\mu^{\bf 1}/\sqrt{3}, d=\m^{\bf 2}/\sqrt{3}, e=\m^{\bf 3'}/\sqrt{6}$.
Since $M_\nu$ is 2-3 symmetric and magic, the tri-bimaximal mixing pattern is guaranteed \cite{LAM0}.

So far we have ignored the right-handed leptons. They must be introduced
to implement a local dynamics, but there is more than one
way to do so. For example, if the right-handed charged leptons, denoted by $c_R$, belongs to {\bf 3},
then the Hamiltonian is once again given by \eq{H}, with $c^\dagger$ replaced by $c_R^\dagger$. The subsequent $S_4$
analyzes are identical, so  we have three isodoublet Higgs coupled to the charged leptons, in
representations ${\bf 1, 3, 3'}$, with the $S_4$-singlet identified with the SM Higgs, and in addition,
three isotriplet Higgs in ${\bf 1,2,3'}$ coupled to the Majorana neutrinos. Other dynamical schemes and the allowed Yukawa
potentials will be discussed in a separate publication later.

In conclusion, we have shown that $S_4$, and groups containing it, are the only finite horizontal symmetries
capable of reproducing
tri-bimaximal mixing of neutrinos without tuning the Yukawa coupling constants. These constants are used
exclusively to fit the fermionic masses.

I am grateful to James Bjorken, Ernest Ma, John McKay, and Maxim Pospelov for helpful discussions.


\begin{thebibliography}{9}
\bibitem{FN} The group $3.S_4$ was denoted in Ref.~\cite{LAM} as $H(12,3)$ and its order was mistaken to be 216. Moreover,
the group $H(6,3)$ was mistaken to have an order of 54; its actual order is 18.
\end{thebibliography}
\end{document}